\documentclass[twocolumn,american,english,nofootinbib]{revtex4-1}
\usepackage[T1]{fontenc}
\usepackage[latin9]{inputenc}
\usepackage{geometry}
\usepackage{color}
\usepackage{babel}
\usepackage{verbatim}
\usepackage{amssymb}
\usepackage{graphicx}
\usepackage{esint}
\usepackage[unicode=true,pdfusetitle,
 bookmarks=true,bookmarksnumbered=false,bookmarksopen=false,
 breaklinks=false,pdfborder={0 0 1},backref=false,colorlinks=true]
 {hyperref}
\hypersetup{
 linkcolor=green}
\usepackage{breakurl}

\makeatletter
\PassOptionsToPackage{hyphens}{url}\usepackage{hyperref}

\makeatother

\begin{document}

\title{Analytical nonlinear collisional dynamics of near-threshold eigenmodes
}

\author{V. N. Duarte}
\email{vduarte@pppl.gov}

\address{Princeton Plasma Physics Laboratory, Princeton University, Princeton,
NJ, 08543, USA}

\author{N. N. Gorelenkov}

\address{Princeton Plasma Physics Laboratory, Princeton University, Princeton,
NJ, 08543, USA}
\selectlanguage{american}%
\begin{abstract}
A closed-form analytical solution is found for the nonlinear dynamics
of isolated, near-threshold waves in the presence of strong scattering.
The proposed solution can be useful in verifying codes across several
disciplines, including Alfvénic instabilities and thermal plasma turbulence
in fusion plasmas and studies of viscous shear flows in fluid dynamics,
as well as a rapid means for predicting and analyzing experimental
outcome.
\end{abstract}
\maketitle
\selectlanguage{english}%
The obtention of reliable bounds for the nonlinear instability of
waves is an outstanding problem in kinetic systems of fusion interest
\citep{Gorelenkov2014,ChenZoncaRevModPhys.2016}. The burning plasma
sustainment in ITER imposes severe constraints on the amount of fast
ions ejected through their resonant interaction with Alfvénic waves
\citep{JacquinotITERExpertGroup1999}. Therefore, procedures to anticipate
the nonlinear evolution of waves destabilized by the sub-population
of highly energetic particles are needed for establishing limits for
wave growth in ITER as well as in present tokamaks. In this letter,
we derive an analytical expression for nonlinear wave evolution in
the presence of strong scattering that can be a rapid means for experimental
prediction and interpretation, as well as for the verification of
codes. 

The nonlinear dynamics of a non-overlapping wave near marginal stability
has been found to be \foreignlanguage{american}{governed by a universal}\footnote{The same equation can be recovered for the evolution of a mode in
a turbulent plasma under a geometric optics approximation, i.e., when
the turbulent modes can be treated as quasi-particles \citep{MendoncaGalvaoPPCF2014}.
A time-delayed, cubic equation of the same structure was also found
in studies of critical layers in shear fluid flows \citep{Hickernell1984}.}\foreignlanguage{american}{ time-delayed, integro-differential cubic
equation} which, in the presence of diffusive processes, reads \citep{BerkPRL1996,BreizmanPoP1997}
\begin{equation}
\begin{array}{c}
\frac{dA(t)}{dt}=A(t)-\frac{1}{2}\int d\Gamma\mathcal{H}\left\{ \int_{0}^{t/2}dzz^{2}A(t-z)\times\right.\\
\left.\times\int_{0}^{t-2z}dye^{-\hat{\nu}_{eff}^{3}z^{2}\left(2z/3+y\right)}A(t-z-y)A^{*}(t-2z-y)\right\} 
\end{array}\label{eq:FullCubicEq}
\end{equation}
where $\hat{\nu}_{eff}$ represents the effective scattering frequency
$\nu_{eff}$ normalized with $\gamma_{L}-\gamma_{d}$ ($\gamma_{L}$
is the linear growth rate in the absence of damping and $\gamma_{d}$
is the sum of a wave background damping rates due to several mechanisms).
Time is also normalized with $\gamma_{L}-\gamma_{d}$. $\hat{\nu}_{eff}$
is an effective frequency due a combination of stochastic processes
experienced by the resonant population, e.g., collisional pitch-angle
scattering, collisionless turbulent scattering and diffusion due to
RF heating waves. The normalized amplitude is $A=\omega_{b}^{2}\gamma_{L}^{1/2}\left(\gamma_{L}-\gamma_{d}\right)^{-5/2}$,
where $\omega_{b}$ is the bounce (or trapping) frequency of the most
deeply trapped resonant particles\footnote{For a simplified bump-on-tail electrostatic case, $\omega_{b}$ is
given by $\sqrt{eEk/m}$ with $e$, $k$ and $m$ being the resonant
particle electric charge, the absolute value of the wave number vector
and the resonant particle mass. For a more realistic toroidal configuration,
$\omega_{b}$ is given by eq. 9 of \citep{BerkPPR1997}. We note that
if our results are to be compared with the ones of Ref. \citep{BerkPPR1997},
our amplitude would need to be divided by a factor $\sqrt{2}$, since
that reference used a slightly modified normalization.}. $d\Gamma$ is a phase-space volume element and $\mathcal{H}$ is
a phase-space weighting defined in \citep{BerkPPR1997,DuartePoP2017}.

Previous numerical analysis for Alfvénic modes in DIII-D, NSTX and
TFTR \citep{DuarteAxivPRL,DuartePoP2017} have shown that the phase
average, over multiple mode resonance surfaces, leads to typical effective
collisional scattering frequency of order $10^{3}s^{-1}$ to $10^{4}s^{-1}$.
Anomalous scattering \citep{LangFu2011} as well as diffusion due
to radiofrequency heating \citep{BergkvistNF2005} contribute to increase
the effective scattering rate. The net growth rate is typically of
order of up to a percent of the wave frequency (the frequecy of toroidicity-induced
and reversed-shear Alfvénic eigenmodes is typically of order $10^{5}s^{-1}$).
Therefore, regimes with $\hat{\nu}_{eff}\gg1$ are relevant for experiments,
especially when the modes are close to threshold and when diffusive
mechanisms, in addition to collisions, are taken into consideration\footnote{In the context of \citep{MendoncaGalvaoPPCF2014}, this limit is equivalent
to very high damping rates of turbulent modes while in the context
of \citep{Hickernell1984} it translates into highly viscous shear
flows.}.

For large scattering frequency, memory effects are easily destroyed
as resonant particles receive frequent random kicks, and only the
very recent history dictates the wave dynamics. For $\hat{\nu}_{eff}\gg1$,
the integral kernel makes the nonlinear term be zero at all times
except when both $z$ and $y$ are close to zero. For very small $y$
and $z$, the kernel of Eq. (\ref{eq:FullCubicEq}) changes much faster
than the arguments of the amplitudes in the cubic term and the term
in the curly brackets can be written as $\frac{A(t)\left|A(t)\right|^{2}}{\hat{\nu}_{eff}^{3}}\int_{0}^{t/2}dz\left[e^{-(2/3)\hat{\nu}_{eff}^{3}z^{3}}-e^{-\hat{\nu}_{eff}^{3}z^{2}\left(3t-4z\right)/3}\right]$.
The argument of the first exponential approaches zero faster than
the one of the second exponential, therefore it is the term that gives
the most important contribution. By redefining the variable of integration
as $x=\hat{\nu}_{eff}z$, the resulting integral can be written as
$\frac{1}{\hat{\nu}_{eff}}\int_{0}^{\infty}dxe^{-(2/3)x^{3}}=\frac{1}{3\hat{\nu}_{eff}}\left(\frac{3}{2}\right)^{1/3}\Gamma\left(\frac{1}{3}\right)$.
We can then seek an analytical solution of the resulting equation

\begin{equation}
\frac{dA(t)}{dt}=A(t)-bA(t)\left|A(t)\right|^{2}\label{eq:ResultingCubicEq}
\end{equation}
by dividing it by $A(t)$ and defining an auxiliary variable $u=\log A$.
Assuming $A(t)\in\mathbb{R}$, a closed-form result is\footnote{In terms of the trapping frequency, the solution is $\omega_{b}(t)=\omega_{b}(0)e^{t/2}/\left[1-c\omega_{b}^{4}(0)\left(1-e^{2t}\right)\right]^{1/4},$
where $c=\left[\gamma_{L,0}/(\gamma_{L,0}-\gamma_{d})\right]\Gamma\left(1/3\right)\left(3/2\right)^{1/3}<\nu_{eff}^{-4}>/6$.
In this expression the time variable $t$ is the actual time multiplied
by $\gamma_{L}-\gamma_{d}$. The average over the resonance surfaces
is defined by $<...>=\int d\Gamma Q.../\int d\Gamma Q$, where $d\Gamma$
is an element of phase space and $Q=\left|e\mathbf{v}\cdot{\bf E}\right|^{2}\left.\partial F/\partial\mathcal{E}\right|_{\mathcal{\mathcal{E}}'}\delta(\Omega-\omega)$,
as defined in \citep{Gorelenkov1999Saturation}}
\begin{equation}
A(t)=\frac{A(0)e^{t}}{\sqrt{1-bA^{2}(0)\left(1-e^{2t}\right)}}\label{eq:AnSol}
\end{equation}
where $A(0)$ is the initial amplitude and $b\equiv\int d\Gamma\mathcal{H}\frac{\Gamma\left(1/3\right)}{6\hat{\nu}_{eff}^{4}}\left(\frac{3}{2}\right)^{1/3}$.
Eq. (\ref{eq:AnSol}) is consistent with its expected asymptotic behaviors
since (i) for $t\rightarrow0$, when the cubic term is unimportant,
the mode grows linearly, i.e., $A(t)=A(0)e^{t}$ provided that $bA^{2}(0)\ll1$
and (ii) for $t\rightarrow\infty$, the saturation level is $A_{sat}=\pm1/\sqrt{b}\simeq\pm1.4/\sqrt{<\nu_{eff}^{-4}>}$
(the sign depends on whether $A(0)$ is positive or negative). Using
the amplitude normalization adopted for Eq. (\ref{eq:FullCubicEq})
we find that, under a bump-on-tail simplification%
, this correponds to the saturation level $\omega_{b,sat}\simeq\pm1.18\left(1-\frac{\gamma_{d}}{\gamma_{L}}\right)^{1/4}\nu_{eff}$,
which agrees with the one previously reported in \citep{BerkPPR1997,PetviashviliThesis}.
To the best of our knowledge, Eq. (\ref{eq:AnSol}) is the first analytical
solution for the mode amplitude evolution, from a seed level up to
saturation, in the presence of collisions.\footnote{An explosive solution \citep{BerkPRL1996} for the cubic equation
(\ref{eq:FullCubicEq}) has been obtained for the situation in which
the linear term is disregarded and the kernel can be replaced by the
unity. The latter signals the breakdown of the theory validity.} 

The associated nonlinear growth rate $\gamma_{NL}\left(t\right)$
can be calculated from $A\left(t\right)=A\left(0\right)\exp\left[\int_{0}^{t}\frac{\gamma_{NL}\left(t'\right)-\gamma_{d}}{\gamma_{L}-\gamma_{d}}dt'\right]$,
which gives
\begin{equation}
\frac{\gamma_{NL}\left(t\right)-\gamma_{d}}{\gamma_{L}-\gamma_{d}}=\frac{1-bA^{2}(0)}{1-bA^{2}(0)\left(1-e^{2t}\right)}.\label{eq:gamma_NL}
\end{equation}

\begin{figure}
\includegraphics[scale=0.35]{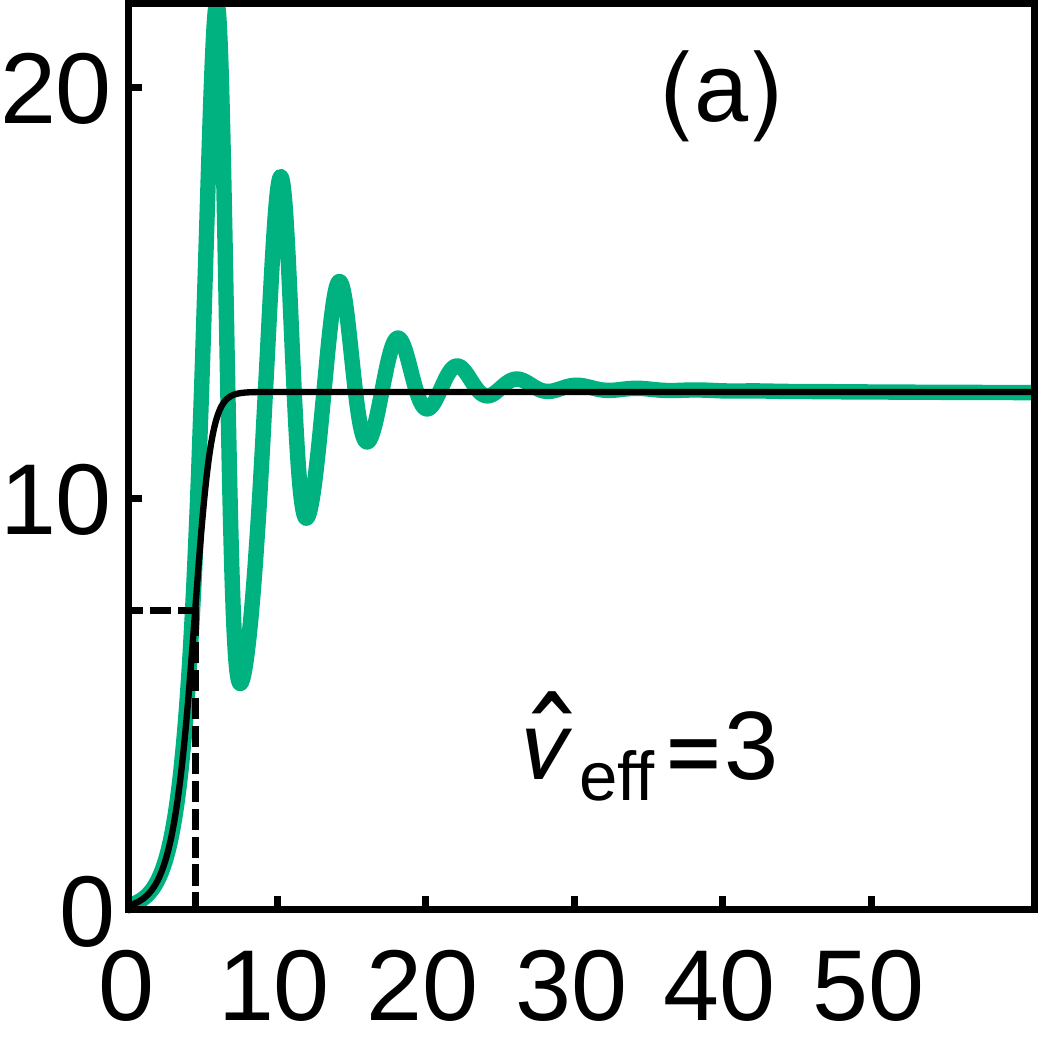} \includegraphics[scale=0.35]{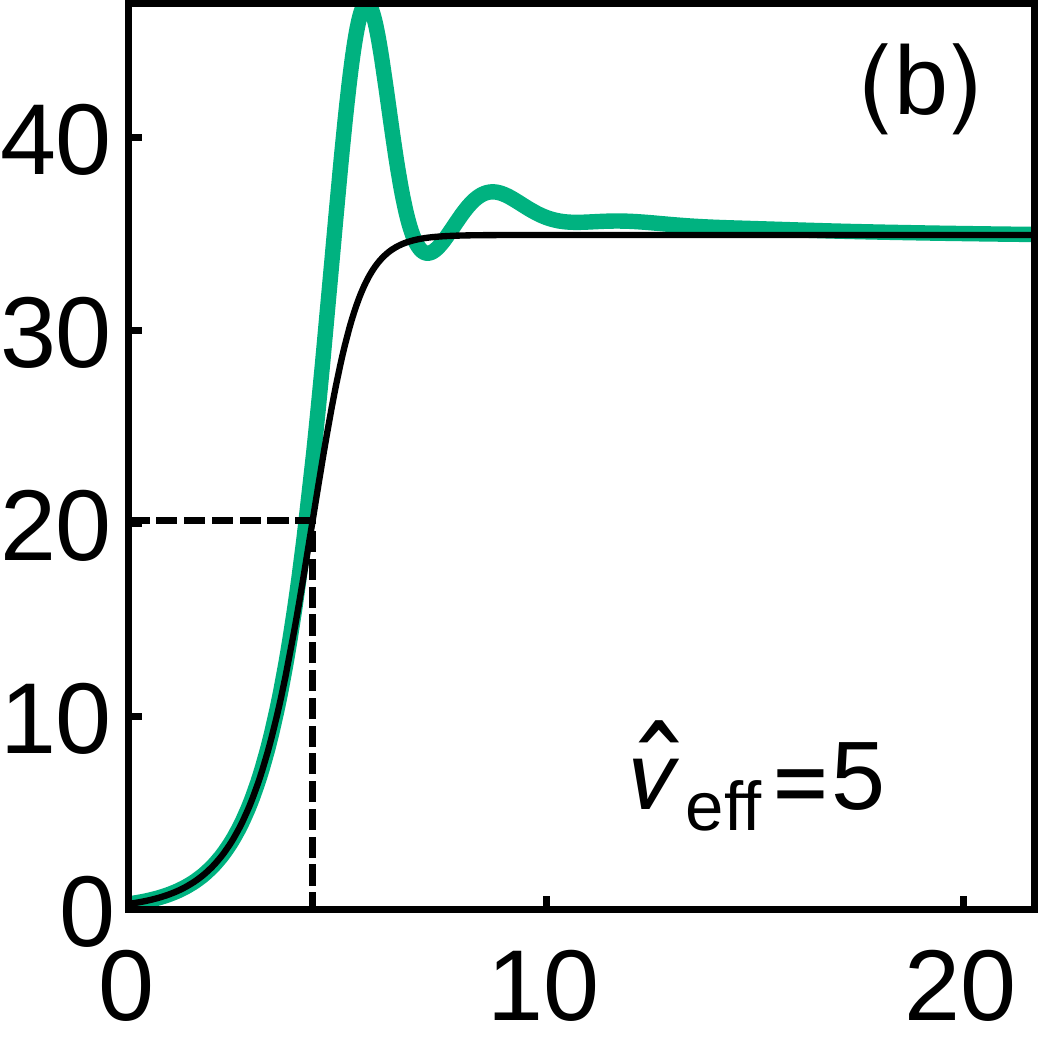}

\includegraphics[scale=0.35]{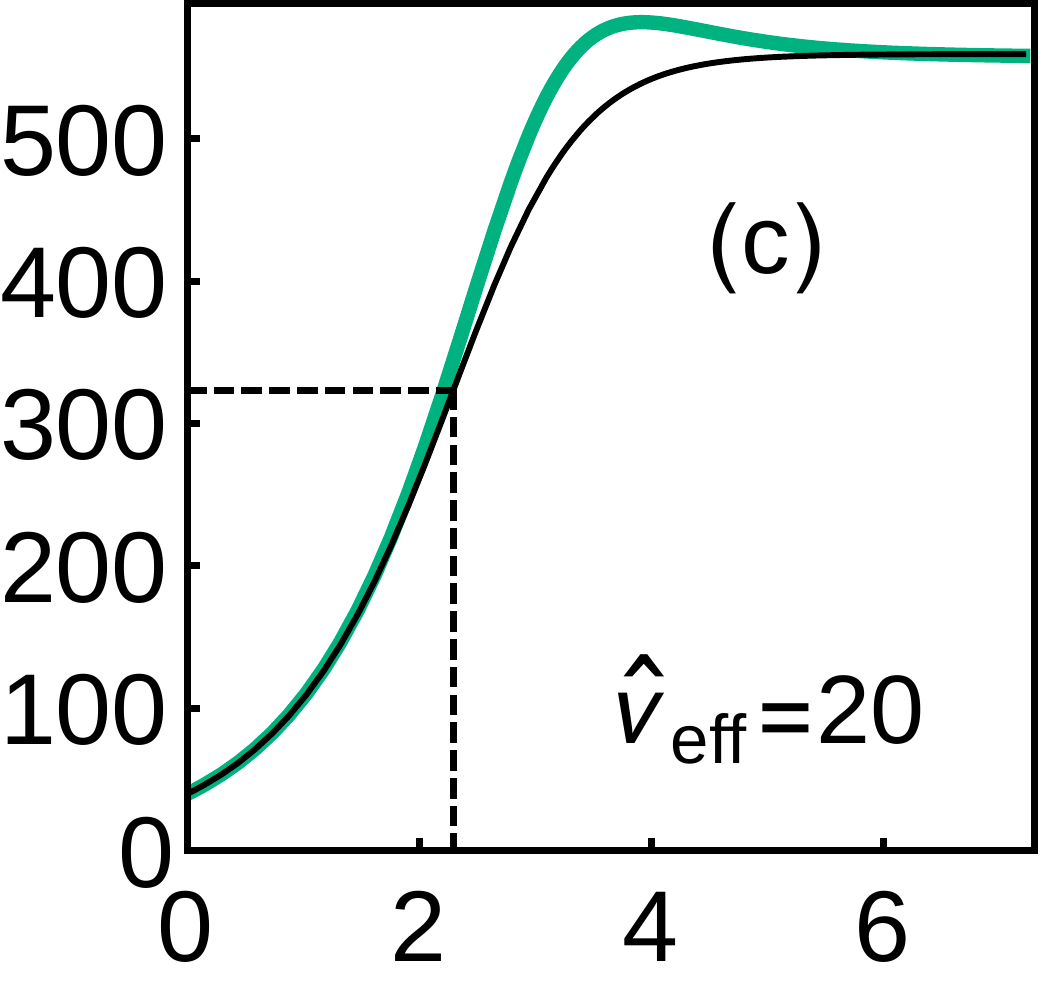} \includegraphics[scale=0.35]{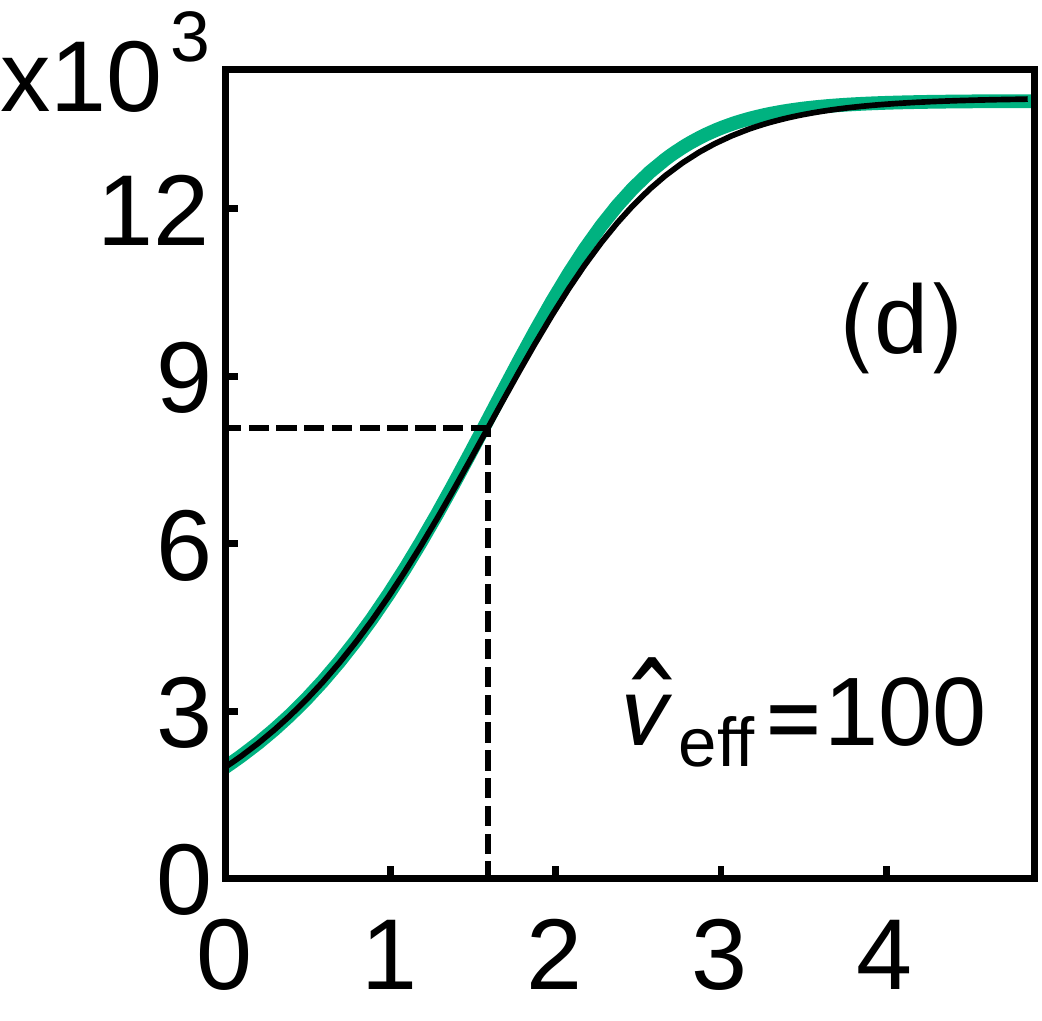}

\caption{\label{fig:A-versus-time} Mode amplitude $A$ versus time $t$ (normalized
with $\gamma_{L}-\gamma_{d}$) for (a) $\hat{\nu}_{eff}=3$, (b) $\hat{\nu}_{eff}=5$,
(c) $\hat{\nu}_{eff}=20$ and (d) $\hat{\nu}_{eff}=100$. In green
is the numerical solution of the full cubic equation (\ref{eq:FullCubicEq})
and in black is the analytical solution (\ref{eq:AnSol}). The dashed
lines indicate the characteristic inflection time for (\ref{eq:AnSol}),
which can vary depending on the choice for $A(0)$ but always happens
at $A_{sat}/\sqrt{3}$. }
\end{figure}

For experimental purposes, it can be useful to anticipate the timescale
for mode saturation, as a function of $\hat{\nu}_{eff}$ and the initial
amplitude $A(0)$. For that purpose, one can gain insights by analyzing
the inflection time point of the solution (\ref{eq:AnSol}), which
is 
\begin{equation}
t_{infl}=\frac{1}{2}\log\left[\frac{1-bA^{2}(0)}{2bA^{2}(0)}\right]\label{eq:InflTime}
\end{equation}
 and corresponds to a characteristic amplitude of $A(t_{infl})=A_{sat}/\sqrt{3}$.
The inflection is indicated on Fig. \ref{fig:A-versus-time}.

In Fig. (\ref{fig:A-versus-time}), we compare the solution for $\hat{\nu}_{eff}\gg1$,
Eq. (\ref{eq:AnSol}), with the full time-delayed cubic equation,
Eq. (\ref{eq:FullCubicEq}), for different values of $\hat{\nu}_{eff}$.
We observe that Eq. (\ref{eq:AnSol}) describes the trace of the wave
amplitude reasonably well \textcolor{black}{for $\hat{\nu}_{eff}\gtrsim2$,
which is when the full cubic equation admits a steady solution \citep{BreizmanPoP1997,HeeterPRL2000}.}\textcolor{red}{{}
}\textcolor{black}{The assumption of high }$\hat{\nu}_{eff}$ used
to derive the analytical solution\textcolor{black}{{} therefore turns
out to be less restrictive than anticipated. In fact, }$\hat{\nu}_{eff}$
simply needs to be\textcolor{black}{{} high enough to ensure steady
saturation, i.e., to prevent the emergence of wave chirping as well
as other higher-order nonlinear bifurcations.}

The existence of a steady solution is always allowed in Eq. (\ref{eq:ResultingCubicEq})
since the linear term can in principle balance the cubic term. The
stability of solution (\ref{eq:AnSol}) can be addressed via eigenvalue
analysis by substituting in Eq. (\ref{eq:ResultingCubicEq}) a perturbed
solution in the form $A_{sat}+\delta Ae^{\left(\lambda_{R}+i\lambda_{I}\right)t}$,
with $\lambda_{R},\lambda_{I}\in\mathbb{R}$. The result is $\lambda_{R}=-2$
and $\lambda_{I}=0$, which means that the saturated solution is intrinsically
stable: any linear perturbation will exponentially asymptote to the
saturation level, without the possibility of oscillations, which are
suppressed by strong scattering processes.

We note that if the collisional scattering kernel of eq. (\ref{eq:FullCubicEq}),
$e^{-\hat{\nu}_{eff}^{3}z^{2}\left(2z/3+y\right)}$, were substituted
by a Krook-type kernel $e^{-\hat{\nu}_{K}\left(2z+y\right)}$ ($\hat{\nu}_{K}$
is the Krook collisional frequency normalized with $\gamma_{L}-\gamma_{d}$),
then solutions of the same type of eq. (\ref{eq:AnSol}), (\ref{eq:gamma_NL})
and (\ref{eq:InflTime}) are admitted, with the transformation $b\mapsto\int\frac{d\Gamma\mathcal{H}}{8\hat{\nu}_{K}^{4}}$.
For the Krook case, the saturation level implied by the analytical
solution is $A_{sat}=2\sqrt{2}\hat{\nu}_{K}^{2}$, in agreement with
\citep{BerkPRL1996}.

\begin{figure}[b]
\begin{centering}
\includegraphics[scale=0.35]{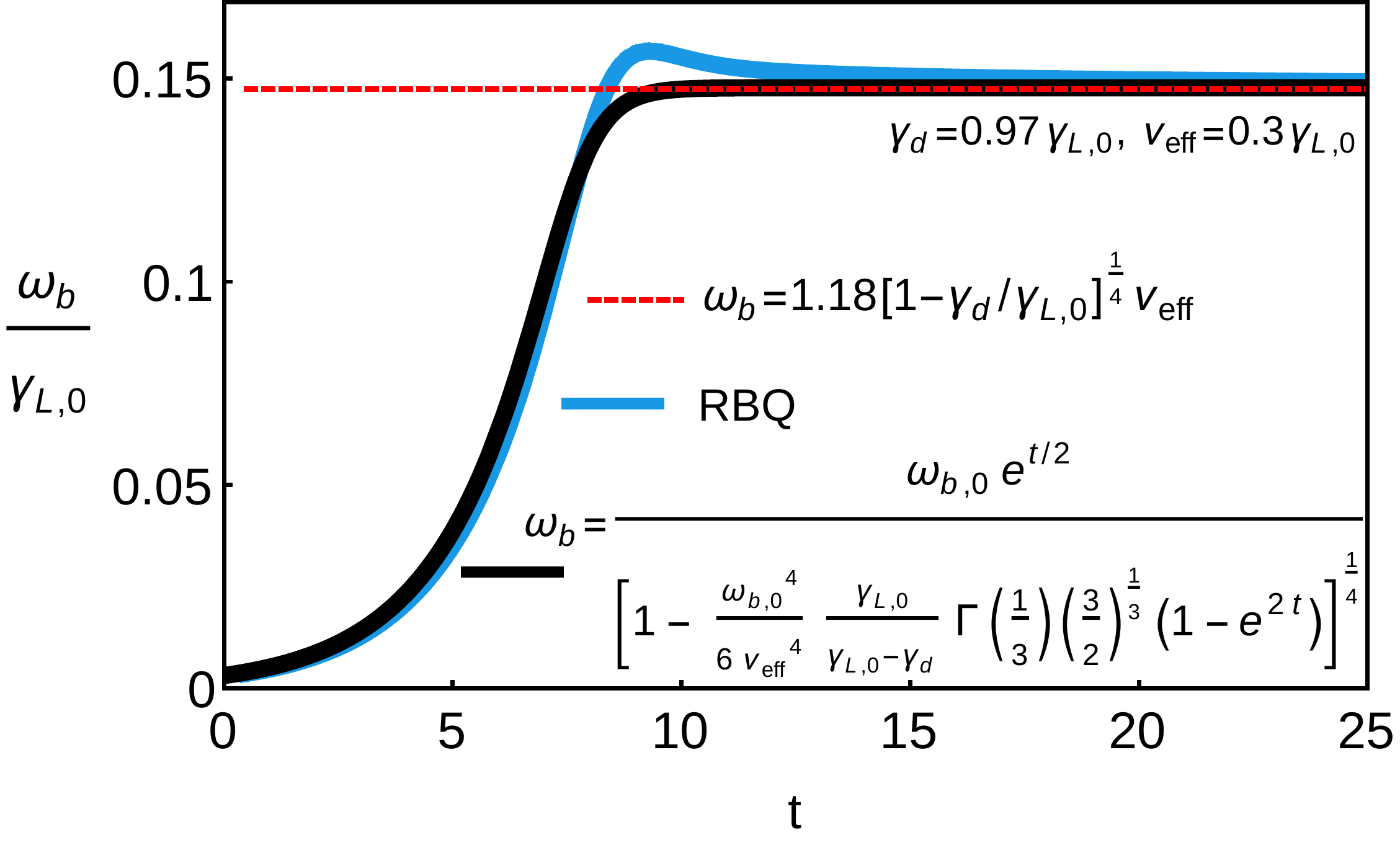}
\par\end{centering}
\caption{\label{fig:LBQ} Comparison between the resonance broadened quasilinear
(RBQ) model \citep{Berk1995LBQ,GhantousPoP2014,GorelenkovDuarteNF2018}
in its bump-on-tail formulation (blue) and the analytical solution
(\ref{eq:AnSol}) (black). The expected saturation level near marginal
stability \citep{BreizmanPoP1997} is shown by the red dashed line.
The parameters used in the simulation are $\gamma_{d}=0.97\gamma_{L,0}$
and $\nu_{eff}=0.3\gamma_{L,0}$. The broadened resonance frequency
$\Delta\Omega=(\pi/2)(1.18)^{4}\gamma_{d}/\gamma_{L,0}$ used in RBQ
ensures that the expected saturation level is achieved.}
\end{figure}

Eqs. (\ref{eq:AnSol}) and (\ref{eq:gamma_NL}) can be used as a verification
for codes, e.g., quasilinear \citep{GorelenkovDuarteNF2018,LiuHirvijokiPRL2018},
gyrofluid \citep{SpongCarrerasPoP1994}, gyrokinetic \citep{LinScience1998,LAUBER2007Ligka,BassWaltzPoP2010,ZarzosoPoP2012,wilkie_abel_highcock_dorland_2015,BassWaltzPoP2017,BiancalaniJPP2017,SlabyNF2018,ColePoP2018},
hybrid (gyro-)kinetic/MHD \citep{FuParkPRL1995,BriguglioPoP1995,TodoMEGAPoP1998,LangFuPoP2010,WangXinNJP2016Saturation,WangXinPoP2016Saturation,ChenYangPoP2018,BriguglioNF2018Saturation},
kinetic \citep{Lesur2010EspectrDetermination,Lilley2010,WoodsDuarteNF2018,Li_NonlSat_PoP2018}
and guiding-center following \citep{PINCHES1998,ChenYangPoP1999,Gorelenkov1999Saturation,ZhouMuni_Roscoe_PPCF2016,WhitePPCF2016Saturation}
simulations for the situation in which the amplitude of a marginally
unstable wave evolves towards a quasi-steady satuaration. Another
possibility to explore the analytical solution \ref{eq:AnSol} is
to compute the distribution function folding within the cubic equation
framework, as recently numerically demonstrated \citep{SanzNF2018}.
A high scattering frequency used in this work destroys phase-space
correlations and therefore prevents the emergence of highly nonlinear
scenarios, such as wave chirping and avalanching. Quasilinear theory
employs a similar reasoning since it neglects the ballistic fast-oscillating
term in its derivation, thereby also not capturing fully nonlinear
wave behavior. An example of the comparison between Eq. \ref{eq:AnSol}
and the RBQ code \citep{GorelenkovDuarteNF2018} is shown in Fig.
\ref{fig:LBQ}, which show fair agreement for regions of parameters
where RBQ does not admit intermittent solutions.

If collisionality is moderate, we note that an amplitude overshoot
occurs following the linear phase, as can be seen from Fig. \ref{fig:A-versus-time}(a).
This can lead to instantaneous wide resonance islands (the resonance
width is roughly proportional to $\omega_{b}$ \citep{MengNF2018}
and therefore proportional to $\sqrt{A}$). The overshoot can be several
times the saturated amplitude, as shown in \citep{ZhouMuni_Roscoe_PPCF2016}.
This may lead to instantaneous overlap of distinct resonances and
invalidate/breaks down the analysis within the cubic equation framework.
Therefore, for purposes of code verification, the expression \ref{eq:AnSol}
applies when collisions are high enough to ensure a monotonic saturation,
in addition to the near threshold regime. As a final remark, we point
out that higher-order nonlinear effects not considered in this work,
such as MHD nonlinearities \citep{ZoncaPRL1995} and wave-wave coupling
\citep{HahmChenPRL1995,QiuPRL2018} can establish further bounds on
the saturation level.
\begin{acknowledgments}
This work was supported by the US Department of Energy (DOE) under
contract DE-AC02-09CH11466. The authors thank V. L. Quito and H. L.
Berk for several discussions.
\end{acknowledgments}

\bibliographystyle{apsrev4-1}
\phantomsection\addcontentsline{toc}{section}{\refname}

\end{document}